\documentclass[12pt,reqno]{amsart}

\usepackage{a4}
\usepackage{amsmath, amssymb}              
\usepackage{graphicx}                      

\swapnumbers                    
\theoremstyle{plain}            
\newtheorem{theorem}{Theorem}[section]
\newtheorem*{theorem*}{Theorem}

\newtheorem{proposition}[theorem]{Proposition}

\theoremstyle{definition}       

\theoremstyle{remark}           
\newtheorem{remark}[theorem]{Remark}

\begin{document}

\newcommand{\R}{\mathbb{R}}            
\newcommand{\C}{\mathbb{C}}            
\newcommand{\Z}{\mathbb{Z}}            
\newcommand{\N}{\mathbb{N}}            
\newcommand{\eps}{\varepsilon}
\newcommand{\la}{\lambda}
\newcommand{\HH}{\mathcal H}
\newcommand{\QQ}{\mathcal Q}
\newcommand{\va}{\vec a}
\newcommand{\ii}{\mathrm i}

\newcommand{\Cci}[1]{C_c^\infty(#1)}   
\newcommand{\lnorm}[1]{\left|\!\left|{#1}\right|\!\right|}

\title{Spectral Gaps for Periodic Elliptic Operators 
   with High Contrast: an Overview}

\author{Rainer Hempel}

\address{Institut f\"ur Analysis, Technische Universit\"at Braunschweig\\
  Pockelsstra{\ss}e 14, 38106 Braunschweig, Germany}
\email{r.hempel@tu-bs.de}

\author{Olaf Post}

\address{Institut f\"ur Reine und Angewandte Mathematik\\
  RWTH Aachen, Templergraben 55, 52062 Aachen, Germany}
\email{post@iram.rwth-aachen.de}  

\date{21.10.2001}

\begin{abstract}
  We discuss the band-gap structure and the integrated density of states for
  periodic elliptic operators in the Hilbert space $L_2(\R^m)$, for $m \ge 2$.
  We specifically consider situations where high contrast in the coefficients
  leads to weak coupling between the period cells. Weak coupling of periodic
  systems frequently produces spectral gaps or spectral concentration.

  Our examples include Schr\"odinger operators, elliptic operators in
  divergence form, Laplace-Beltrami-operators, Schr\"odinger and Pauli
  operators with periodic magnetic fields. There are corresponding
  applications in heat and wave propagation, quantum mechanics, and photonic
  crystals.
\end{abstract}

\maketitle

\section*{Introduction}
We consider periodic elliptic partial differential operators of the second
order with high contrast in (some of) the coefficients. The classes of
operators discussed here include Schr\"odinger operators, magnetic
Schr\"odinger and Pauli operators, the acoustic operator, Laplace-Beltrami
operators, and Maxwell operators. We are mainly interested in the question
whether high contrast in the coefficients leads to weak coupling between the
period cells and to spectral concentration and/or spectral gaps. For most of
the paper, we assume a simple geometry like the one shown in Figure~1, for
which the strongest results are obtained.  We include some remarks on more
general situations, like sponges in dimensions $3$ and higher. Our main theme
is to show that simple topological assumptions lead to spectral consequences
for a wide class of---rather different---elliptic operators. We do not discuss
the question of absolute continuity here.

The paper is organized as follows. Section~1 contains some basic definitions,
assumptions and background results concerning monotone convergence of
quadratic forms. In Section 2, we briefly report on some ``classical'' results
on the convergence of Schr\"odinger operators with high barriers. Similar
results for strong magnetic barriers in the Schr\"odinger and Pauli case are
discussed in Section 3. Section 4 is devoted to divergence type operators
$-\nabla \cdot (1 + \la \chi_\Omega) \nabla$, where $\Omega$ is a periodic,
open subset of $\R^m$, with $m \ge 2$. There is some superficial
similarity between divergence type operators and Laplace-Beltrami operators.
The existence of spectral gaps for periodic Laplace-Beltrami operators with a
conformal factor in the metric tensor is discussed in Section 5.  We also
refer to constructions of a purely ``geometric'' nature that yield
Laplace-Beltrami-operators with gaps.  We conclude in Section 6 with some
references to the work of A.~Figotin, P.~Kuchment and others on the Maxwell
operator and applications to photonic crystals.

\section{Preliminaries and basic assumptions}
\label{sec:prelim}

Throughout the present paper, we will always assume that $\Omega \subset \R^m$
is open and periodic w.r.t.\ some discrete lattice $\Gamma$; for most purposes
it is enough to consider $\Gamma = \Z^m$. $Q$ denotes a fundamental period
cell of $\Gamma$; we may take $Q = (0,1]^m$ if $\Gamma = \Z^m$. The complement
of $\Omega$ is denoted as $M$, a closed and periodic set.  We will study
two-component media with material parameter $\la =1$ on $M$, and $ \la \gg 1$
on $\Omega$.  The strongest results are obtained in the ``model situation'' of
Figure~1 where $\Omega$ contains the boundary of the period cell $Q$.
\begin{figure}[t]
  \begin{center}
    \includegraphics{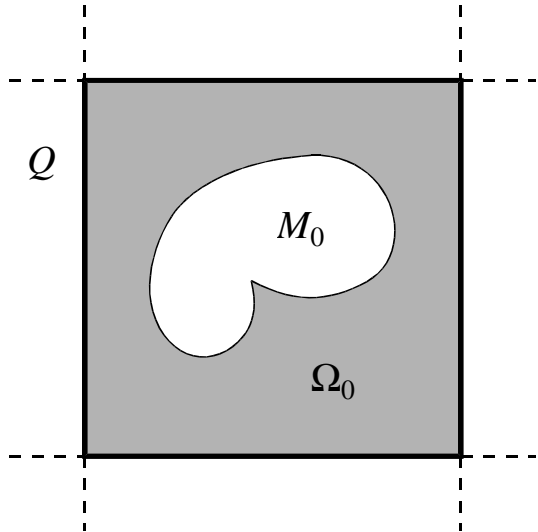}
    \caption{The fundamental period cell $Q$ and the set $M_0=Q \cap
      M$. \label{fig:per.cell}}
  \end{center}
\end{figure}

Our basic Hilbert spaces are 
$\HH = L_2(\R^m)$ and $\HH^1(\R^m)$, the Sobolev space of $L_2$-functions 
that have weak derivatives in $L_2$. 

All our operators are semi-bounded, self-adjoint operators $T$ acting in 
some $L_2$-space; they are obtained from some closed, semi-bounded quadratic 
form ${\mathfrak t}$ via the usual representation theorem ([K, RS-I]) and satisfy 
$D(T) \subset D({\mathfrak t})$ and the relation
$$
    {\mathfrak t}[u,v] =  \left<{Tu},{v}\right>,
                \qquad u \in D(T), \quad v \in D({\mathfrak t}). 
    \eqno{(1.1)}
$$
As usual, we say $T_2 \ge T_1$ iff the associated quadratic forms satisfy
$D({\mathfrak t}_2) \subset D({\mathfrak t}_1)$ and ${\mathfrak t}_2[u,u] \ge
{\mathfrak t}_1[u,u]$, for all $u \in D({\mathfrak t}_2)$. We also write 
${\mathfrak t}[u] := {\mathfrak t}[u,u]$ in the sequel. 

For a sequence of non-negative, self-adjoint operators $T_n$ in $\HH$ we say
that $T_n$ converges to a limit $T_\infty$ in the \emph{strong resolvent
  sense} iff
$$
           (T_n + I)^{-1}f \to  (T_\infty + I)^{-1}f, \qquad f \in \HH;  
         \eqno{(1.2)}
$$  
for convergence in the \emph{norm resolvent sense} we require uniform
convergence for all $ f \in \HH$ s.th.\ $\lnorm{f} \le 1$.

Most of our results hinge upon the following convergence theorem for quadratic
forms ([RS-I], [S]):   

\begin{theorem}
  \label{thm:1.1}
  Let $(T_n)$ be a monotonic sequence of non-negative, self-adjoint operators.
  Then there exists a self-adjoint operator $T_\infty \ge 0$ s.th.  $ T_n \to
  T_\infty$ in strong resolvent sense.
\end{theorem}

For an \emph{increasing} sequence of quadratic forms $0 \le {\mathfrak t}_n \le
{\mathfrak t}_{n+1}$ with domains $\QQ_n$ the limiting form domain is given
by $\QQ_\infty = \{ u \in \cap_{n=1}^\infty \QQ_n ; \sup {\mathfrak
  t}_n[u] < \infty\} $.  Here it may happen that $T_\infty$ acts in a strictly
smaller Hilbert space than the $T_n$. In this case, $(T_\infty + I)^{-1}$ has
to be complemented by the zero operator on a suitable subspace of $\HH$.  For a
\emph{decreasing} sequence it is shown in [RS-I, S] that there exists a largest
closable form that is smaller than ${\mathfrak t}_n$, for all $n$.

In a periodic situation, strong resolvent convergence often implies
convergence of spectral densities. For the definition of the \emph{integrated
  density of states} (i.d.s.) we refer to [RS-IV], [PF] or [DIT]. In our
context, the i.d.s.\ is a monotonically increasing function $F : \R \to \R$
that tends to zero at $-\infty$. Such functions have at most a countable
number of discontinuities.  The associated Borel measure $\mu$ is then called
the \emph{density of states measure}.  We will say that a sequence $(\mu_n)$
of density of states measures converges weakly to $\mu_\infty$ if $F_n(t) \to
F_\infty(t)$ at all points of continuity $t$ of $F_\infty$. The proof of the
subsequent proposition is elementary if one uses the definition of [RS-IV] for
the i.d.s.

\begin{proposition} 
 \label{prop:1.2}  
 In addition to the assumptions of Theorem 1.1, suppose that the operators
 $T_n$ are periodic (with respect to the the same lattice $\Gamma$),
 and that a (finitely-valued) i.d.s.~$F_n$ exist for $T_n$
 as well as for $T_\infty$.  Then the corresponding measures $\mu_n$ converge
 weakly to $\mu_\infty$.
\end{proposition} 
It is well-known that \emph{norm resolvent convergence} implies spectral
convergence (cf.~[RS-I; Ch.~VIII]) on any compact interval of the real line.
In particular, we have the following simple consequence for the existence of
spectral gaps: suppose $T_n \to T_\infty$ in norm-resolvent sense. If
$T_\infty$ has a gap around $E \in \R$, then the $T_n$ will also have a gap
around $E$, for large $n$. More precisely, suppose that $(a,b) \cap
\sigma(T_\infty) = \emptyset$ and let $\eps > 0$. Then $(a+\eps, b-\eps) \cap
\sigma(T_n) = \emptyset$, for all $n \ge n_\eps$.

\section{Schr\"odinger operators}
\label{sec:schroe.op}

As a warm-up, we illustrate the above convergence theorems by some simple
results on Schr\"odinger operators of the form $-\Delta + \la V(x)$, where $V$
is non-negative.  It has been a ``folk theorem'' for quite some time that
$-\Delta + \la \chi_B$ converges in strong resolvent sense to the Dirichlet
Laplacian on the complement $\R^m \setminus B$, under suitable
assumptions on $B$.

Suppose $V(x) \ge 0$ is a periodic, real-valued (continuous or measurable and
bounded) potential s.th.\ $\{x \in \R^m \mid V(x) \ne 0\} = \Omega$. We let
$$
  H_\lambda = -\Delta + \la V(x), \qquad \la \ge 0, \eqno{(2.1)}
$$
with associated quadratic forms ${\mathfrak h}_\la[u] = \lnorm{\nabla u}^2 +
\la \int V |u|^2$, for $u \in \HH^1$.

To describe the limit, we will need here (and in the following sections) the
\emph{Dirichlet Laplacian} on the closed set $M = \R^m \setminus \Omega$,
denoted as $-\Delta_M$ ([Hed, HZ]).  The operator $-\Delta_M$ is
constructed from the quadratic form $\lnorm{\nabla u}^2$ with domain
$$
    {\tilde\HH}^1_0(M) := \{ u \in \HH^1(\R^m) \mid u(x) = 0 
\hbox{\rm\ a.e.\ in\ } M^C\}. 
  \eqno{(2.3)}
$$
 For sufficiently regular boundary of $M$,
 $-\Delta_M$ agrees with the usual Dirichlet Laplacian. 
 In the general case, it is shown in [HZ] that there exists a Borel set $M^*$, 
 $M^\circ \subset M^* \subset M$, such that  $-\Delta_M$ acts as a
 self-adjoint  operator in the Hilbert-space $L_2(M^*)$.
  It is now easy to see that 
$$
   \{ u \in \HH^1(\R^m) \mid \sup_{\la \ge 0} {\mathfrak h}_\la[u] < \infty \}
   =  {\tilde\HH}^1_0(M), 
   \eqno{(2.4)}
$$
and so Theorem~1.1 implies that  
$H_\la$ converges in strong resolvent sense to $-\Delta_M$.
(To be more precise, the resolvent of $-\Delta_M$ acts in $L_2(M^*)$ and 
 has to be complemented by
 the zero operator on $L_2(\R^m \setminus M^*)$ in Eqn.~(1.1).) Up to 
this point, periodicity was not required. If we now assume that $\Omega$ is 
periodic, then for all operators involved 
the i.d.s.\ is well-defined, and we may also conclude that the i.d.s.\ converges. 

As was observed in [HH-1], one can actually do better and upgrade to 
norm resolvent convergence. The proof combines uniform exponential decay estimates  
for the resolvent kernels with Schur's Lemma and local compactness
of $-\Delta$. 

\smallskip

The above convergence results do not require $\Omega$ to be connected.  In our
``model case'' of Figure~1 where $\Omega$ contains a neighborhood of $\partial
Q$, $-\Delta_M$ is a countable direct sum of copies of $-\Delta_{M \cap Q}$.
Therefore, the spectrum of $-\Delta_M$ is a discrete set with each point in
the spectrum an eigenvalue of infinite multiplicity. We therefore see that the
spectrum of $H_\la$ concentrates at a discrete set of points, and an
arbitrarily large number of spectral gaps opens up, as $\la \to \infty$.

\section{Magnetic Schr\"odinger and Pauli operators}
\label{sec:mag.op}

Here we briefly mention some results of [HH-1,2], [HN], [B-1,2] on strongly
coupled periodic magnetic fields.

\smallskip
(1) Let $\va \in C^1(\R^m, \R^m)$ be a magnetic vector potential and 
 $B = d\va$ the associated magnetic field. 
We assume that $B$ is periodic w.r.t.\ the lattice $\Gamma$,  
and that $B(x) = 0$ for all $x \in M$,
 while $B(x) \ne 0$ for all $x \in \Omega$. Note that periodicity of 
 $\va$ is \emph{not} required. 
The magnetic Schr\"odinger operator is then given as [CFKS]
$$
   H(\la{\vec a}) = ( -\ii \nabla - \la{\vec a})^2, 
   \eqno{(3.1)}
$$
and obtained via  the closure of the forms 
$$
 \left< H(\la{\vec a}) u,u \right> = 
   \lnorm{(-\ii\nabla - \la {\vec a}(x))u}^2, \qquad u \in \Cci{\R^m}.
$$
Evidently, there is \emph{no} monotonicity w.r.t.\ $\lambda$.  However, in any
open subset $U$ of $\Omega$ where a fixed entry $B_{ij}(x)$ of the magnetic
tensor is bounded away from zero, the Avron-Herbst-Simon bound [AHS] gives a
\emph{local} lower bound for the quadratic form, expressing the barrier-like
effect of a strong magnetic field. If the vector potential $\va$ vanishes on
$M$, the above results for $-\Delta + \mu \chi_U$, $\mu \to \infty$, can be
combined with the Feynman-Kac-It\^o-formula to obtain strong resolvent
convergence of $H(\la\va)$ to $ -\Delta_M$, as $\la \to \infty$ ([HH-1]). For
given magnetic field $B$, it depends on the topology of the set $\Omega$ and
on flux conditions whether a gauge exists s.th.\ $d\va = B$ and $\va(x) = 0$
a.e.\ on $M$.

Under the above conditions, we also 
obtain an upgrade to norm-resolvent convergence, as before, and 
get the same spectral consequences as for 
 $-\Delta + \la \chi_\Omega$. 

\smallskip 

(2) The situation where $\va(x)$ is allowed to have non-zero values on $M$ has
been analyzed in detail in [HN], using some basic homology theory and de
Rham's Theorem. It can be shown that, under suitable assumptions, the norm
difference of the resolvents of $H(\la\va)$ and $H_M(\la\va)$ tends to zero,
as $\la \to \infty$, where $H_M(\la\va) = ( -\ii \nabla - \la{\vec a})^2$,
acting in $L_2(M^*)$ with Dirichlet boundary conditions.  As a consequence,
one can conclude that the spectrum of $H(\la\va)$ approaches a periodic or a
quasi-periodic function of $\la$, as $\la \to \infty$.  The precise result
involves the flux of $B$ through the connected components of
$(\overline{M^\circ})^C$, the complement of the closure of the interior of
$M$.

\vskip1ex 

(3) For the {\bf Pauli operator} in $\R^2$, we have the pair of operators 
$$
   H_\pm(\la\va) = H(\la\va) \mp B,  
  \eqno{(3.2)}
$$
both acting in $L_2({\mathbb R}^2)$. 
Here supersymmetry (cf.~[CFKS]) implies 
$$
                 \sigma(H_+(\la\va)) \setminus \{0\} = 
\sigma(H_-(\la\va)) \setminus \{0\}.
  \eqno{(3.3)}
$$
Note that the results in (1) and (2) remain unchanged if we include an 
electric background potential $V$ (e.g., $V$ bounded and measurable), while 
supersymmetry is destroyed upon addition of an electric potential. 

Under suitable assumptions, one can show ([B-1,2]) that the 
 operators $H_\pm(\la\va)$ have 
a common gap: in the simplest case, we assume $B \ge 0$.  
A more delicate result requires that 
 $B$ is positive in a neighborhood of $\partial Q$
 and that there exists a periodic gauge s.th.\ $d\va = B$. (By the divergence 
theorem, this gauge condition is equivalent to the \emph{flux condition} 
$\int_Q B(x) dx = 0$).

\section{Periodic divergence type operators}
\label{sec:per.div.op}

Here we report on some results of [HL-1,2] concerning periodic two-component
media with high contrast. In the following we always assume $m \ge 2$. The
operators dicussed below occur in acoustics, heat conduction, and propagation
of electro-magnetic waves in photonic crystals.

The quadratic form is given by the \emph{Dirichlet integral}
$$
   {\mathfrak t}_\lambda[u] := \int ( 1 + \lambda \chi_\Omega(x)) 
          | \nabla u(x)|^2 dx, 
   \qquad u \in \mathcal{H}^1, \quad \la \ge 0;
   \eqno{(4.1)} 
$$
this defines a  monotonically increasing family of forms. The associated 
self-adjoint operators $T_\la$ can formally be written as  
 $T_\la = - \nabla \cdot (1 + \la \chi_\Omega) \nabla$. 

It is clear that only such functions $u$ from the form domain survive 
taking $\la \to \infty$ that are constant on each connected component of 
$\Omega$. If $\Omega$ is connected, this implies that only such $u$'s 
will remain in the limiting form domain $\QQ_\infty$ that are identically zero 
on $\Omega$. We have: 
\begin{theorem} {\rm [HL-1]}
  \label{thm:4.1}
  Suppose $\Omega$ is open, periodic, and connected. Then $T_\la$
  converges to the Dirichlet-Laplacian $-\Delta_M$ on $M = \R^m \setminus
  \Omega$ in strong resolvent sense. Furthermore, the associated density of
  states measures converge weakly.
\end{theorem} 
The standard Floquet-decomposition gives a more detailed picture.  Let
$T_\la^{(\theta)}$ denote the operator $-\nabla \cdot ( 1 + \la \chi_\Omega)
\nabla$, acting in $L_2(Q)$, with $\theta$-periodic boundary conditions on
$\partial Q$, for $\theta \in (-\pi, \pi]^m$; cf.~[RS-IV, Ku]. Then Theorem 1.1
can be applied in each fiber, i.e., for each fixed $\theta$. As each
$T_\la^{(\theta)}$ has compact resolvent, we even obtain norm resolvent
convergence in the fibers. The limiting operators $T_\infty^{(\theta)}$ have
form domains $\QQ_\infty^{(\theta)}$ consisting of those functions $u \in
\HH^1(Q)$ that are constant in each connected component of $\Omega \cap Q$ and
satisfy $\theta$-periodic boundary conditions. If $\Omega$ is connected and
intersects each face of the period cell $Q$, it follows that such $u$ must
vanish on $\Omega \cap Q$, provided $\theta \ne (0,\ldots,0)$. Pursuing this
observation in detail, one obtains

\begin{theorem} {\rm ([HL-1])}
 \label{thm:4.2}  
 Suppose $\Omega \subset \R^m$ is open, periodic, connected and contains
 $\partial Q$. We also assume that $M^*$ (defined in Section~2) is not of
 measure zero. Let $0 < \delta_1 \le \delta_2 \le \ldots$ denote the
 eigenvalues of $-\Delta_{M \cap Q}$.  We then have:

 $(i)$ The first spectral band of $T_\la$ converges to the interval
 $[0,\delta_1]$.
 
 $(ii)$ For $\la$ large, $T_\la$ has a spectral gap following the first band.
 
 $(iii)$ The i.d.s.\ measure of $T_\la$ concentrates at the set $\{\delta_k ;
 k \in \N \}$, as $\la \to \infty$.
\end{theorem}
It follows from $(i)$ that there 
is \emph{no} norm resolvent convergence of the family $(T_\la)$ (The 
only possible candidate for the limit is $-\Delta_M$ which has no 
spectrum below $\delta_1 > 0$. But norm resolvent convergence implies 
spectral convergence.) 
 
For smoothly bounded $M$ a much more detailed picture of the process 
of spectral concentration is given in [F]. 
\begin{remark} 
  \label{rem:4.3} 
  More general results can be found in [HL]. First of all, without additional
  effort one can handle elliptic operators of the type $-\nabla \cdot
  ({\mathbf a}(x) + \la {\mathbf b}(x)) \nabla$, under suitable assumptions on
  the (symmetric) matrices ${\mathbf a}(x)$ and ${\mathbf b}(x)$, like
  $a_{ij}, b_{ij} \in L_\infty(\R^m)$, ${\mathbf a}(x)$ uniformly positive
  definite, ${\mathbf b}(x) = {\mathbf 0}$ on $M$, and positive definite for
  $x \in \Omega$.
  
  Second, [HL] also gives criteria for the existence of higher gaps: if the
  eigenspace of the $k$-th Dirichlet eigenvalue $\delta_k$ of $-\Delta_{M \cap
    Q}$ contains an eigenfunction $d_k$ s.th.\ $\int d_k \ne 0$ then
  $\sigma(T_\la)$ will have a gap for $\la$ large that converges to
  $(\delta_k, \nu_k)$, for some $\nu_k > \delta_k$.
\end{remark}

\begin{remark} 
  \label{rem:4.4}
  There are natural and simple situations where one is led to consider period
  cells that are more general than $(0,1]^m$.  In fact, all of the above works
  if $\Gamma$ is a discrete sublattice of $\R^m$ that spans $\R^m$ and if $Q$
  is a contractible subset of $\R^m$ with piecewise smooth boundary
  tessalating $\R^m$ via the translations $\gamma \in \Gamma$.  Here we think,
  in particular, of the class of results that depend on the assumption that
  $\Omega$ contains the boundary of the period cell.
\end{remark} 

\begin{remark} 
  \label{rem:4.5}
  In $\R^m$, $m \ge 3$, it is possible to have both $\Omega$ and $M$
  connected. A \emph{sponge} is a typical example (albeit, in reality, not
  periodic but random!).  Here our analysis would yield that, for high
  contrast, the i.d.s.\ is approximately given by the i.d.s.\ of $T_\infty$.
  The latter can be further analyzed by Floquet-decomposition if we assume
  that $\Omega$ intersects each face of $Q$ (think of the case where $M$ is
  connected from one cell to the neighboring cells through thin filaments
  only).
\end{remark} 

\begin{remark} 
  \label{rem:4.6}
  It is natural to ask what happens for a \emph{decreasing} sequence of
  operators like
  $$
    S_\la = - \nabla \cdot \frac 1 {1 + \la \chi_M} \nabla, \qquad 
    0 \le \la \uparrow \infty. 
    \eqno{(4.2)} 
  $$
  The associated quadratic forms ${\mathfrak s}_\la$ are defined by
  ${\mathfrak s}_\la[u] = \int (1 + \la \chi_M)^{-1} |\nabla u|^2 dx$ on
  $\HH^1(\R^m)$.  Evidently, the largest form that is smaller than ${\mathfrak
    s}_\la$, for all $\la \ge 0$, has domain $\HH^1(\R^m)$ and takes the
  values $\int_\Omega |\nabla u|^2 dx$. Let us call this form ${\mathfrak
    s}_\infty$.  It is easy to see that ${\mathfrak s}_\infty$ is closable.
  Let $S_\infty$ denote the self-adjoint operator associated with the closure
  of ${\mathfrak s}_\infty$.  It follows from Simon's theorem [RS-I;
  Thm.~S.~16] that the operators $S_\la$ converge to $S_\infty$ in strong
  resolvent sense.
  
  Let us assume, for the moment, that any $u \in \HH^1(\Omega)$ can be
  extended to a function in $\HH^1(\R^m)$ and that $\Cci{M^\circ}$ is dense in
  $L_2(M)$.  Then it is not difficult to show that the closure of the form
  ${\mathfrak s}_\infty$ has domain $\HH^1(\Omega) \oplus L_2(M)$, and takes
  the values ${\mathfrak s}_\infty[u] = \int_\Omega |\nabla u(x)|^2 dx$.
  Under these assumptions $S_\infty$ is the direct sum of the Neumann
  Laplacian on $\Omega$ and the zero operator on $L_2(M)$. In particular, the
  i.d.s.\ of $S_\infty$ would be infinite on the positive real line, and it
  seems rather hopeless to extract any useful spectral information on $S_\la$.
  (Of course, the r\^oles of $M$ and $\Omega$ are interchangeable here if the
  boundary of $M$ is smooth enough.)
\end{remark} 

\begin{remark} 
  \label{rem:4.7}
  In the theory of photonic crystals, one is led to study the
  ``high-contrast/thin-wall limit'' of the acoustic operator in $\R^2$.
  In typical applications, the optically dense medium occupies a connected,
  periodic set $M$ consisting of thin walls, cf.~[FK-1,2]. In our notation, we
  would then consider the limit of $T_\la = -\nabla ( 1 + \la \chi_\Omega)
  \nabla$, as $\la \to \infty$.  Note that now $\Omega$ is \emph{not}
  connected.  Our results yield a strong resolvent limit and a
  Floquet-decomposition analogous to what we have discussed above. The
  limiting operators $T_\infty^{(\theta)}$ in the fibers now have a form
  domain consisting of functions that are constant on $\Omega \cap Q$, where
  the constant value will be non-zero, in general. Therefore, it is much
  harder to determine the spectrum of the $T_\infty^{(\theta)}$ than in the
  case of Eqn.~(4.1). If $\Omega \cap Q$ exhausts ``most of'' $Q$ (thin wall
  asymptotics), then the analysis can proceed again [FK-1,2], [KK].
\end{remark} 

\begin{remark}
  \label{rem:4.8}
  Random Media and the convergence of the i.d.s.\ for the operators of
  Eqn.~(4.1) with a random geometry have been studied by K.~Lienau [L].  Since
  we need a connected $\Omega$ for our approach to work well, the ergodic
  process that generates the random set $M$ has to respect this property. Here
  various methods from stochastic geometry ([SKM]) can be employed.
\end{remark}

\section{Periodic Laplace-Beltrami operators}
\label{sec:per.lap.bel}

In this section we report on some related results on periodic Laplace-Beltrami
operators, cf.~[P-1,2].  Let $\mathcal M$ be a non-compact Riemannian manifold
of dimension $d\ge 2$.  We call $\mathcal M$ a \emph{periodic} or
\emph{covering manifold} iff $\Gamma=\Z^m$ acts properly discontinuously and
isometrically on $\mathcal{M}$.  Under these conditions, the quotient
$\mathcal M / \Gamma$ is again a Riemannian manifold which we assume to be
compact.

We call $Q$ a \emph{period cell} or \emph{fundamental domain} of $\mathcal M$
iff $Q$ is open, $Q$ does not intersect any other translate $\gamma + Q$
except for $\gamma=0$, and the union of all translates of $\overline Q$ covers
the whole manifold $\mathcal M$. Note that we assume here that $Q$ is an open
set. One can apply Floquet decomposition in the same way as before; here,
$\vartheta$-periodicity means that $u(\gamma + x)=e^{i \vartheta \cdot \gamma}
u(x)$ for all $x \in \overline Q$ and $\gamma \in \Gamma$ such that $\gamma +
x \in \overline Q$ (plus an analogous condition for the first derivatives).

Denote by $(g_{ij})$ the metric tensor of $\mathcal M$ in some chart.  We
consider conformal deformations of the given periodic metric, i.e., we set
$$
           g_{ij}(x;\la) = \varrho^2_\la(x) \, g_{ij}(x).
$$ 
Here, $\varrho_\la$ is a strictly positive, periodic, smooth function
converging pointwise to the indicator function of a closed periodic set $M$ as
$\la \to \infty$. In particular, we assume that
\smallskip

{\leftskip1em 

 (1) $M_0:=M \cap Q$ is compactly contained in $Q$ with smooth $\partial M$;
\vskip.5ex 

(2) $\varrho_\la(x)=1$ for all $x \in M$;  
\vskip.5ex

(3) $\varrho_\la (x) = \frac 1 \la$, for all $x \in \Omega$ with dist$(x,M)
 \ge \la^{-m}$.

}
\smallskip

To simplify the notation, we restrict ourselves to the conformally flat case,
i.e., we assume that $\mathcal{M}=\R^m$ with the Euclidean metric tensor
$g_{ij}(x)=\delta_{ij}$, defined on $\R^m$.  We assume that the boundary
$\partial M$ has a tubular neighborhood where we can introduce local
coordinates $(r,y)$ given by the distance $r$ from $\partial M$ and $y \in
\partial M$. For technical reasons we need the additional assumption that
these coordinates can be extended to $\Omega \cap Q$; e.g., think of a ball
$M_0$ centered in the period cell $Q=(0,1)^m$.  The Laplace-Beltrami-operator
$B_\la$ corresponding to the deformed metric is given by
$$
    B_\la = - \frac 1 {\sqrt{g_\la}}
 \partial_j  \left( g^{ij}(\cdot;\la) \sqrt{g_\la} \right) \partial_i 
$$
(using the summation convention),  
where $g_\la=\det g_{ij}(\cdot,\la)=\rho_\la^{2m}$  denotes the square 
of the  volume density of the deformed metric tensor. 
 This operator is defined as a non-negative, 
self-adjoint operator via the quadratic form 
$$
   {\mathfrak b}_\la[u] :=
     \int g^{ij}(\cdot,\la) 
          \partial_i u \, \partial_j \overline u \sqrt{g_\la} \, dx=
     \int |\nabla u|^2 \, \varrho_\la^{m-2} \, dx,
   \qquad u \in \mathcal{H}^1(\R^m)
   \eqno{(5.1)} 
$$
in the Hilbert space $\HH_\la:=L_2(\R^m)$ with inner product $\int u
{\overline v} \sqrt{g_\la} dx$. Note that, in general, both the Dirichlet
integral \emph{and} the volume form depend on the parameter $\la$.  The
fundamental difference to the case of divergence type operators lies in the
dependence of the inner product on $\la$.  In particular, for $m \ge 3$, there
is no monotonicity.

Floquet-decomposition implies that it is enough to study $\vartheta$-periodic
b.v.p.'s on $Q$.  Furthermore, Dirichlet-Neumann-bracketing yields enclosures
of bands by Neumann and Dirichlet eigenvalues $E_k^{N/D}(\la)$ (ordered by
min-max and repeated according to multiplicities).  Similarly, denote by
$E_k^N$ the Neumann eigenvalues of $M_0=M \cap Q$.

Since the norm of the Hilbert space depends on $\lambda$, we cannot apply
Theorem~1.1. Instead, the convergence of the eigenvalues on $Q$ to the Neumann
eigenvalues on $M_0$ is obtained directly through an application of the
min-max-principle and a careful comparison of the corresponding Raleigh
quotients.

In particular, we show that $E_k^N$ is (approximately) a lower bound for
$E_k^N(\lambda)$ (using the restriction $u |_{M_0}$ as transition operator
from $\HH^1(Q)$ to $\HH^1(M_0)$). In the same way, we prove that $E_k^N$ is
also an approximate upper bound for $E_k^D(\lambda)$ provided $m \ge 3$ (using
an extension operator from $\HH^1(M_0)$ to $\HH_0^1(Q)$, the $\HH^1$-closure
of $\Cci Q$, as transition operator). Therefore, we have
\begin{theorem} {\rm ([P-1,2])}
  \label{thm:5.1}
  Let $m \ge 3$. Then $\sigma(B_\la)$ converges to the spectrum of the Neumann
  Laplacian on $M$, as $\la \to \infty$. In particular, gaps open up, as $\la
  \to \infty$.
\end{theorem}

\begin{remark} 
  \label{rem:5.2}
  The case $m = 2$ is special. Here, only the inner product depends on $\la$.
  Inverting the r\^ole of the norm and the quadratic form yields a limiting
  operators in each fiber which, however, depend on $\theta$. In the special
  case of a $2$-dimensional cylinder one can nevertheless prove the existence
  of spectral gaps by direct calculations (separation of variables),
  cf.~[P-1,2].
\end{remark} 

\begin{remark} 
  \label{rem:5.3}
  In contrast to the Euclidean case, we may choose the dimension $d$ of the
  manifold $\mathcal M$ independently of the dimension of periodicity $m$:
  think of the surface of a cylinder ($d=2, m=1$) or the surface of a periodic
  jungle gym ($d=2, m=3$). Furthermore, one can allow any Abelian finitely
  generated group $\Gamma$ of infinite order, i.e., products of $\Z$ and
  cyclic groups.
\end{remark}

\begin{remark} 
  \label{rem:5.4}
  In some sense, the case of Laplace-Beltrami operators is the opposite of the
  case of divergence type operators: while the divergence type operator needs
  the factor $(1+\la \chi_\Omega)$ to be \emph{large} on $\Omega$ to produce
  spectral gaps, the Laplace-Beltrami operator needs the factor
  $\varrho_\la^{m-2}$ to be \emph{small} on $\Omega$.
\end{remark} 

\begin{remark}
  \label{rem:5.5}
  Davies and Harrell~[DH] proved the existence of at least one gap in the
  periodic conformally flat case (transforming the conformal Laplacian into a
  corresponding Schr\"odinger operator). This result is a special case of our
  result. Using similar methods as~[DH], Green [G] obtained examples with a
  finite number of gaps in the $2$-dimensional conformally flat case.
  Recently, Yoshitomi~[Y] proved the existence of spectral
  gaps for the (Dirichlet) Laplacian on periodically curved quantum wave
  guides. In [DH, G, Y], the existence of gaps is established by analyzing a
  one-dimensional problem. Here, in contrast, we directly study the
  multi-dimensional problem.
\end{remark} 

\begin{remark} 
  \label{rem:5.6}
  More ``geometric'' constructions of periodic manifolds with spectral gaps
  can be found in [P-1,2], think for example of an infinite
  number of copies of a compact manifold, joined by thin cylinders.  If these
  cylinders are small or thin enough, one again obtains spectral gaps of the
  corresponding Laplace-Beltrami operator. These results are closely related
  to earlier works of Chavel and Feldman~[CF] and Ann\'e~[A].
\end{remark}

\section{Maxwell operators}
\label{sec:maxwell.op}

Another important elliptic operator is the \emph{Maxwell operator} given as
$$
    M  = - \hbox{\rm curl} \frac 1 {\eps(x)} \hbox{\rm curl},  
   \eqno{(6.1)}
$$
acting on soleno{\"\i}dal vector fields in ${\mathbb R}^2$ or in ${\mathbb
  R}^3$, where $\eps(x)$ denotes the dielectricity of the medium.

The existence of gaps for periodic structures is of fundamental importance for
photonic crystals. In many cases, one is satisfied to find intervals of low
spectral density (quasi-gaps). This class of problems has been studied in
depth in a series of papers by A.~Figotin, P.~Kuchment and others (cf., e.g.,
[FK-1,2], [KK]) where a complete analysis for high-contrast and/or thin-wall
asymptotics is given.

\end{document}